\documentclass{elsart}
\usepackage[square,comma]{natbib}
\usepackage{amsmath}

\input{tcilatex}

\input psfig
\begin{document}

\runauthor{W. Rolke} 
\begin{frontmatter}
\title{Confidence Intervals and Upper Bounds for Small Signals in the Presence of Background Noise\thanksref{X}}
\author[wolf]{Wolfgang A. Rolke}
\address[wolf]{Department of Mathematics, University of Puerto Rico - Mayag\"{u}ez, Mayag\"{u}ez, PR 00681, USA, 
\newline Postal Address: PO Box 5959, Mayag\"{u}ez, PR 00681, 
\newline Tel: (787) 851-6212, Email: w\_rolke@rumac.upr.clu.edu}
\author[ang]{Angel M. L\'{o}pez}
\address[ang]{Department of Physics, University of Puerto Rico - Mayag\"{u}ez, Mayag\"{u}ez, PR 00681, USA}
\thanks[X]{This work was supported by the High Energy Physics Office (Grant DE-FG-97ER41045) and the EPSCoR Program (Grant DE-FG-94ER75764) of the US Department of Energy  and the University of Puerto Rico.}

\begin{abstract}
We discuss a new method for setting limits on small signals in the presence of background noise. The method is based on a combination of a two dimensional confidence region and the large sample approximation to the likelihood ratio test statistic.
It automatically quotes upper limits for small signals and two-sided confidence intervals for larger samples. We show that this method gives the correct coverage and also has good power. 
\end{abstract}
\begin{keyword}
Maximum likelihood, profile likelihood, confidence regions, coverage, Monte Carlo, sensitivity
\end{keyword}
\end{frontmatter}
\newpage

\section{Introduction}

Finding confidence intervals or upper limits for small signals has recently
attracted a great deal of attention. The paper by Feldman and Cousins \cite
{Cousins-Feldman} rekindled the interest in this area by providing a unified
approach, that is, it based the choice of quoting an upper limit or a
two-sided confidence interval on the data alone, without the experimenter
having to make this decision. The unified approach uses the Neyman
construction together with an ordering principle based on likelihood ratios.
Subsequent papers such as Giunti \cite{Giunti} and Roe and Woodroofe \cite
{Roe-Woodroofe} gave variations of this method, basing the ordering on other
quantities. Common to all these methods is the need to have a fairly precise
knowledge of the background rate, for example, from Monte Carlo simulations.
Unfortunately, as we will see in section 4, these methods can fail when used
in the presence of background uncertainty. A possible remedy is discussed in
Cousins and Highland \cite{Cousins-Highland} where it is suggested to treat
the background uncertainty as a systematic error. We will instead treat the
background uncertainty as a statistical error and develop a method that is
suitable in this situation. Our method is based on the likelihood ratio test
statistic, together with an adjustment for those situations where there is
very little (or no) signal observed. We will show that this method yields
the correct coverage rate and that it has good power.

The background rate has to be estimated either from the data or through
Monte Carlo. Both of those methods have their strengths and their
weaknesses. Using data requires choosing sidebands, and implicitly makes the
assumption that the density generating the background events is the same in
the signal region as it is in the sidebands. This leads to a quandary: If we
choose a small sideband, this assumption seems more reasonable, but then we
will also see fewer background events and therefore have a higher
statistical uncertainty in the estimate of the background rate. Choosing a
large region might yield higher statistics but makes the assumption of a
linear background more tenuous. An alternative way to estimate the
background rate is by Monte Carlo. One problem with this approach is that a
good Monte Carlo is often difficult to do because we can never be completely
sure that we have modeled all the relevant effects correctly. This is
particularly true when searching for small signals since the efforts to
reduce the background often mean one is probing the tails of distributions
which may be difficult to model. Also, in High Energy Physics today running
a complete Monte Carlo simulation of an experiment can be a formidable task
from a computational point of view, and it might not be possible to run
enough Monte Carlo to effectively eliminate the uncertainty in the
background estimate. For these reasons, it is in many cases not possible to
ignore the uncertainty in the background rate.\ Our method is well equipped
to deal with the uncertainty that comes from limited statistics in\ both
situations, those where the background is estimated from the data as well as
those where the background is estimated using Monte Carlo.

In some cases there are two different sources of background. It turns out
that our method can be extended fairly easily to this situation also,
regardless of the method of estimation used.

\section{A Description of the Method}

In this section we will outline the basic ideas of this new method. We will
need the following notation. Assume that we observe $x$ events in a suitably
chosen signal region and a total of $y$ events in the background region.
Here the background region can be chosen fairly freely and need not be
contiguous. Furthermore, we assume that the ratio of the size of the
background region to the size of the signal region is $\tau $. For example,
if we use two background regions of the same size as the signal region we
get $\tau =2$. Then a probability model for the data is given by 
\begin{equation*}
X\sim Pois(\mu +b),\qquad Y\sim Pois(\tau b)
\end{equation*}
where $\mu $ is the signal rate, $b$ is the background rate and $Pois$ is
the usual Poisson distribution. We can assume $X$ and $Y$ to be independent
and so 
\begin{equation*}
P_{\mu ,b}(X=x,Y=y)=\frac{(\mu +b)^{x}}{x!}e^{-(\mu +b)}\cdot \frac{(\tau
b)^{y}}{y!}e^{-\tau b}
\end{equation*}

\subsection{A Confidence Region for $\protect\mu $ and $b$}

A common technique for constructing confidence regions (or intervals for
one-dimensional parameters) is to find a corresponding hypothesis test and
then to invert the test. We will start with a simultaneous hypothesis test
for $\mu $ and $b$ with the null hypothesis $H_{0}:\mu =\mu _{0},b=b_{0}$.
The steps are as follows:

\begin{enumerate}
\item  List all observations $(u_{i},v_{i}),i=1,..,K$, together with their
probabilities, which are given by 
\begin{equation*}
P_{\mu _{0},b_{0}}(X=u_{i},Y=v_{i})=\frac{(\mu _{0}+b_{0})^{u_{i}}}{u_{i}!}%
e^{-(\mu _{0}+b_{0})}\cdot \frac{(\tau b_{0})^{v_{i}}}{v_{i}!}e^{-\tau b_{0}}
\end{equation*}
Here we use the values $\mu _{0}$ and $b_{0}$ specified in the null
hypothesis. We list only those observations which have a probability above a
certain small threshold. In our algorithm we require the probabilities to be
larger than $10^{-6}$.

\item  Sort all observations from the most likely to the most unlikely.

\item  Find the partial sums from the largest to the $k^{th}$ observation
until you reach $1-\alpha ,$ if the desired level of the test is $\alpha .$

\item  If the observed $(x,y)$ appears in the list of possible observations
before $1-\alpha $ is reached, accept the null hypothesis, otherwise reject
it.
\end{enumerate}

This is in effect a Neyman construction like the one used in Feldman and
Cousins \cite{Cousins-Feldman}, only we use the probabilities as the
ordering quantities. The test simply checks whether or not the observation $%
(x,y)$ is compatible with the rates $\mu _{0}$ and $b_{0}$ specified in the
null hypothesis.

Using the likelihood ratios as the ordering quantity in a way similar to
Feldman and Cousins \cite{Cousins-Feldman} would in principle be superior,
but unfortunately does not yield a viable method here because the list of
''likely'' observations is infinite.

The inversion of the hypothesis test then involves a search through all
pairs $\left( \mu ,b\right) $.\ If a certain pair leads to the acceptance of
the null hypothesis, we add it to the confidence region, otherwise we do
not. As an example we have figure 1 where we show the confidence regions
obtained for three observations. The boundaries of the confidence regions
are somewhat ragged due to the discrete nature of the Poisson random
variable.

\subsection{A Confidence Interval for $\protect\mu $}

Again we will start with a hypothesis test, but this time we will only fix
the signal rate $\mu $. The null hypothesis then becomes $H_{0}:\mu =\mu
_{0} $.

A popular test in Statistics for any kind of hypothesis test is the
likelihood ratio test, which is based on the likelihood ratio test statistic 
$\Lambda $ given in our problem by: 
\begin{equation*}
\Lambda (\mu _{0};x\mathbf{,}y)=\frac{\max \left\{ l(\mu _{0},b;x,y):b\geq
0\right\} }{\max \left\{ l(\mu ,b;x,y):\mu \geq 0,b\geq 0\right\} }
\end{equation*}

Here $l(\mu ,b;x,y)=P_{\mu ,b}(X=x,Y=y)$ is the likelihood function of $\mu $
and $b$ given the observation$(x,y)$. This test statistic can be thought of
as the ratio of the best explanation for the data if $H_{0}$ is true and the
best explanation for the data if no assumption is made on $\mu $. The
denominator is simply the likelihood function evaluated at the usual maximum
likelihood estimator. To find the numerator we have to find the maximum
likelihood estimator of the background rate $b$ assuming that the signal
rate is known to be $\mu _{0}.$

\begin{equation*}
\frac{\partial }{\partial b}\log l(\mu _{0},b;x,y)=\frac{x}{\mu _{0}+b}-1+%
\frac{y}{b}-\tau \doteq 0
\end{equation*}
\begin{equation*}
\widehat{b}=\frac{x+y-(1+\tau )\mu _{0}+\sqrt{\left( x+y-(1+\tau )\mu
_{0}\right) ^{2}+4(1+\tau )y\mu _{0}}}{2(1+\tau )}
\end{equation*}
$l(\mu ,\widehat{b};x,y)$ is called the profile likelihood function of $\mu $%
. The usefulness of the likelihood ratio test statistic lies in the fact
that approximately we have 
\begin{equation*}
-2\log \Lambda (\mu _{0};x,y)\sim \chi ^{2}(d)
\end{equation*}
that is $-2\log \Lambda $ has an approximate Chi-Square distribution with $d$
degrees of freedom, where $d$ is the number of parameters in the model minus
the number of parameters specified in the null hypothesis. Here we have $d=1$%
. Because the profile likelihood $l$ differs from the likelihood ratio $%
\Lambda $ only by a constant independent of $\mu _{0}$, we can concentrate
on the profile likelihood. For more details on the likelihood ratio test
statistic see Casella and Berger \cite{Casella-Berger}. For information on
the profile likelihood see Bartlett \cite{Barlett}, Lawley \cite{Lawley} and
Murphy and Van Der Vaart \cite{Murphy and Van Der Vaart}. In figure 2 we
have the profile likelihood function for the case $x=6$, $y=2$, $\tau =2$.
To find a $(1-\alpha )\cdot 100\%$ confidence interval we start at the
minimum, which of course is at the usual maximum likelihood estimator, and
then move to the left and to the right until the function rises by the $%
\alpha $ percentile of a $\chi ^{2}$ distribution with $1$ degree of freedom.

The method here uses an approximation to the profile likelihood function by
a quadratic function. Unfortunately, in cases where the number of
observations in the signal region is small compared to the number of
background events, the profile likelihood function becomes almost linear and
this approximation does not work. For those cases we will use the following
method:

We overlay the confidence region described previously with the profile
likelihood curve $\left( \mu ,\widehat{b}\right) $. Then we find the
smallest value of $\mu $ that is on this curve but not in the confidence
region. Figure 3 illustrates this method. Clearly, in the case of fewer
observations in the signal region than in the background region only an
upper bound will be quoted. We will use this second method whenever the
profile likelihood function has a positive derivative at $\mu =0$. It turns
out that the limits obtained by these two methods are compatible. Figure 4
shows the upper bound for a number of cases with the limits obtained by the
confidence region method drawn as squares and the limits using the
likelihood ratio method as diamonds. The transition from one method to the
other is quite smooth.

\section{Extensions of the Method}

\subsection{Estimating background from Monte Carlo}

Our method can also be applied when a Monte Carlo with limited statistics is
used to estimate the background rate. Assume we run the Monte Carlo $n$
times and observe a total of $y$ events. In the data we see $x$ events in
the signal region. Then a probability model for this situation is given by: 
\begin{equation*}
X\sim Pois(\mu +b),Y\sim Pois(nb)
\end{equation*}
We notice, of course, that this is actually the same model as the one
previously used, only with an $n$ instead of a $\tau $. Therefore, we can
use our method for this situation without any changes.

\subsection{Include a second background source}

Sometimes there is a second source of background present in the data, this
one characterized by the fact that it only appears in the signal region. An
example is an invariant mass histogram where some of the background comes
from the misidentification of pions with\ muons. Say we run a Monte Carlo $n$
times and observe a total of $z$ events of this type. In the data we have $x$
observations in the signal region and $y$ observations in a suitably chosen
background region. The probability model for this case is given by : 
\begin{equation*}
\begin{array}{c}
X\sim Pois(\mu +b+\eta ),Y\sim Pois(\tau b) \\ 
Z\sim Pois(n\eta )
\end{array}
\end{equation*}
where $\mu $ is the signal rate, $b$ is the rate of the first background
source and $\eta $ is the rate of the second background source. Then: 
\begin{equation*}
P_{\mu ,b,\eta }(X=x,Y=y,Z=z)=\frac{(\mu +b+\eta )^{x}}{x!}e^{-(\mu +b+\eta
)}\frac{(\tau b)^{y}}{y!}e^{-\tau b}\frac{(n\eta )^{z}}{z!}e^{-n\eta }
\end{equation*}
We can extend our method to this situation in a fairly straightforward
manner. First, we have to find the profile likelihood function, which leads
to the following nonlinear system of equations: 
\begin{equation*}
\frac{\partial \log l}{\partial b}=\frac{x}{\mu +b+\eta }-1+\frac{y}{b}-\tau
=0
\end{equation*}
\begin{equation*}
\frac{\partial \log l}{\partial \eta }=\frac{x}{\mu +b+\eta }-1+\frac{z}{%
\eta }-n=0
\end{equation*}
It turns out that $\widehat{\eta }$ can be found as the second largest root
of the cubic equation $ax^{3}+bx^{2}+cx+d=0$ with 
\begin{equation*}
\begin{tabular}{l}
$a=(1+n)(n-\tau )$ \\ 
$b=\left( x+z-(1+n)\mu \right) (\tau -n)-(1+n)(z+y)$ \\ 
$c=xz+(\tau -n)z\mu +z^{2}+yz-(1+n)z\mu $ \\ 
$d=\mu z^{2}$%
\end{tabular}
\end{equation*}
and then $\widehat{b}$ is given by 
\begin{equation*}
\widehat{b}=\frac{x+y-(1+\tau )(\mu +\widehat{\eta })+\sqrt{\left(
x+y-(1+\tau )(\mu +\widehat{\eta })\right) ^{2}+4(1+\tau )y(\mu +\widehat{%
\eta })}}{2(1+\tau )}
\end{equation*}
The same problem as before arises again in this situation: in the case of
few events in the signal region the profile likelihood function is nearly
linear. We can use the same remedy as before by instead finding the
intersection of the boundary of the three-dimensional confidence region in $%
(\mu ,b,\eta )$ space with the profile likelihood curve $(\mu ,\widehat{b},%
\widehat{\eta })$.

\section{Performance of this Method}

In this section we will study the true coverage and the power of this
method. For comparison we will use the unified approach of Feldman and
Cousins \cite{Cousins-Feldman}. Although that method was not designed to
deal with uncertainty in the background, it is the standard method used in
calculating confidence intervals in High Energy Physics at this time
according to the Review of Particle Physics \cite{Particle Data Group}. The
coverage rates shown here were obtained through exact computation, without
approximation. In figure 5 we have the case of one background region of
equal size to the signal region, and finding $90\%$ confidence intervals.
The true background rates are $b=1,3$ and $5$ and the signal rates go from $%
0 $ to $5.$ Clearly our method has much better coverage than Feldman and
Cousins \cite{Cousins-Feldman}; although, due to the approximation used in
the method, the coverage is sometimes slightly worse than the nominal one.
The ragged appearance of the graph is due to the discrete nature of the
Poisson distribution and in general is unavoidable. Figure 6 has the case of
two background regions and a $99\%$ confidence interval. Again the new
method performs very well, whereas Feldman and Cousins \cite{Cousins-Feldman}
does not achieve the nominal coverage.

Next, we illustrate the performance of this method when the background rate
is estimated by Monte Carlo. In figure 7 we compute the coverage rates for
the cases $\mu =0$ and $b=0.5,1.5$ and $2.5$. The Monte Carlo is run $n$
times where we let $n$ range from $1$ to $10$. Our method performs
satisfactorily for all cases. This plot is also interesting because it gives
some insight into the number of runs of the Monte Carlo needed before one
can assume that the background rate is known.

The performance of the extension of our method to the case of two different
background sources discussed in section 3.2 has to be studied using mini
Monte Carlo because the number of possible observations $(x,y,z)$ becomes
quite large. We have run a variety of these mini Monte Carlo studies. In
Figure 8 we show the results for the case $\tau =2$, $n=10$ and a $90\%$
confidence interval. The true coverage rates of our method appear to be in
line with the nominal ones, again with a few cases where the coverage is
slightly worse due to the approximation used in the method. Due to the
discreteness of the Poisson distribution the method is also quite
conservative for many situations.

As Feldman and Cousins \cite{Cousins-Feldman} noted, there has been some
criticism of their limits in cases where an experiment finds fewer events
than are expected from the background rate. For example, an experiment with
an expected background rate of $b=0.5$ and no observed events would quote a
Feldman-Cousins upper limit of $1.94$; whereas an experiment with an
expected background rate of $b=1.5$ and no observed events would quote a
Feldman-Cousins upper limit of $1.33$. Our tables show a similar effect.
Feldman and Cousins \cite{Cousins-Feldman} explain this apparent
inconsistency in some detail, and we are in full agreement with their
reasoning.

It may also come as somewhat of a surprise that our method yields limits
that are sometimes smaller than the limits in Feldman and Cousins \cite
{Cousins-Feldman}.Contrary to the instance discussed in the previous
paragraph, this happens in some cases where more events are observed in the
signal region than in the sidebands. For example, consider the following
situation: say we use one sideband of equal size to the signal region, that
is we have $\tau =1$, and we observe $x=1$ events in the signal region and $%
y=0$ events in the sideband. Then the $90\%$ upper limit using our method is
found to be $3.65$, whereas assuming $b=0$ and using Feldman and Cousins
method gives a $90\%$ upper limit of $4.36$, or about $19\%$ larger. So here
we get a smaller limit despite the fact that we have an additional
uncertainty. This apparent paradox can be explained as follows: Let us
consider the question whether the single event observed in the signal region
is a signal or a background event. Using Feldman and Cousins \cite
{Cousins-Feldman} the answer is clear: having assumed $b=0$ we know that
there is no background, therefore this event has to be from the signal. In
fact, it would be quite reasonable in this situation to actually quote a
lower limit larger than $0$, because we already have proof that $\mu >0$. On
the other hand, in our method $y=0$ does not imply $b=0$, it only means that 
$b$\ is not too large. It is therefore still possible that the event in the
signal region is in fact a background event, and that $\mu =0$. It should
come as no surprise, then, that in those cases we are quoting a smaller
upper limit for $\mu $ than Feldman and Cousins \cite{Cousins-Feldman}.
Notice that in theory it would still be preferable to have absolute
knowledge of the background rate because one could then make an actual
observation of the signal rate rather than having to settle for setting a
limit. In practice it is usually impossible to have such precise knowledge
and, of course, nobody would claim a discovery based on just one event.

\section{Conclusion}

The methods for quoting limits for rare decays, mainly Feldman and Cousins 
\cite{Cousins-Feldman} and its variants, all suffer from the requirement
that the background source be known with a high degree of precision. We have
described a new method based on the likelihood ratio test which treats the
background uncertainty as a statistical error. The performance of the method
is shown to be quite good, with the true coverage rates close to the nominal
ones and good power. The method can be used in situations where the
background rate has been estimated from data sidebands as well as where it
has been obtained from Monte Carlo. The method can also be extended to cases
where a second background source solely appearing in the signal region is
present.

In the appendix we provide tables for the confidence intervals and the
experimental sensitivity as discussed in Feldman and Cousins \cite
{Cousins-Feldman} and in Review of Particle Physics \cite{Particle Data
Group}\ for the cases $\tau =1,2$ and $\alpha =0.9,0.99$. The currently used
algorithm for computing the limits becomes unreliable in some extreme cases,
and in those cases we use NA in the tables. A FORTRAN program for the
computation of the limits for any other case as well as for the extension
discussed in section 4.2. can be obtained by writing to
w\_rolke@rumac.upr.clu.edu.

\newpage

\section{Appendix}

\begin{figure}[ht]
\psfig{file=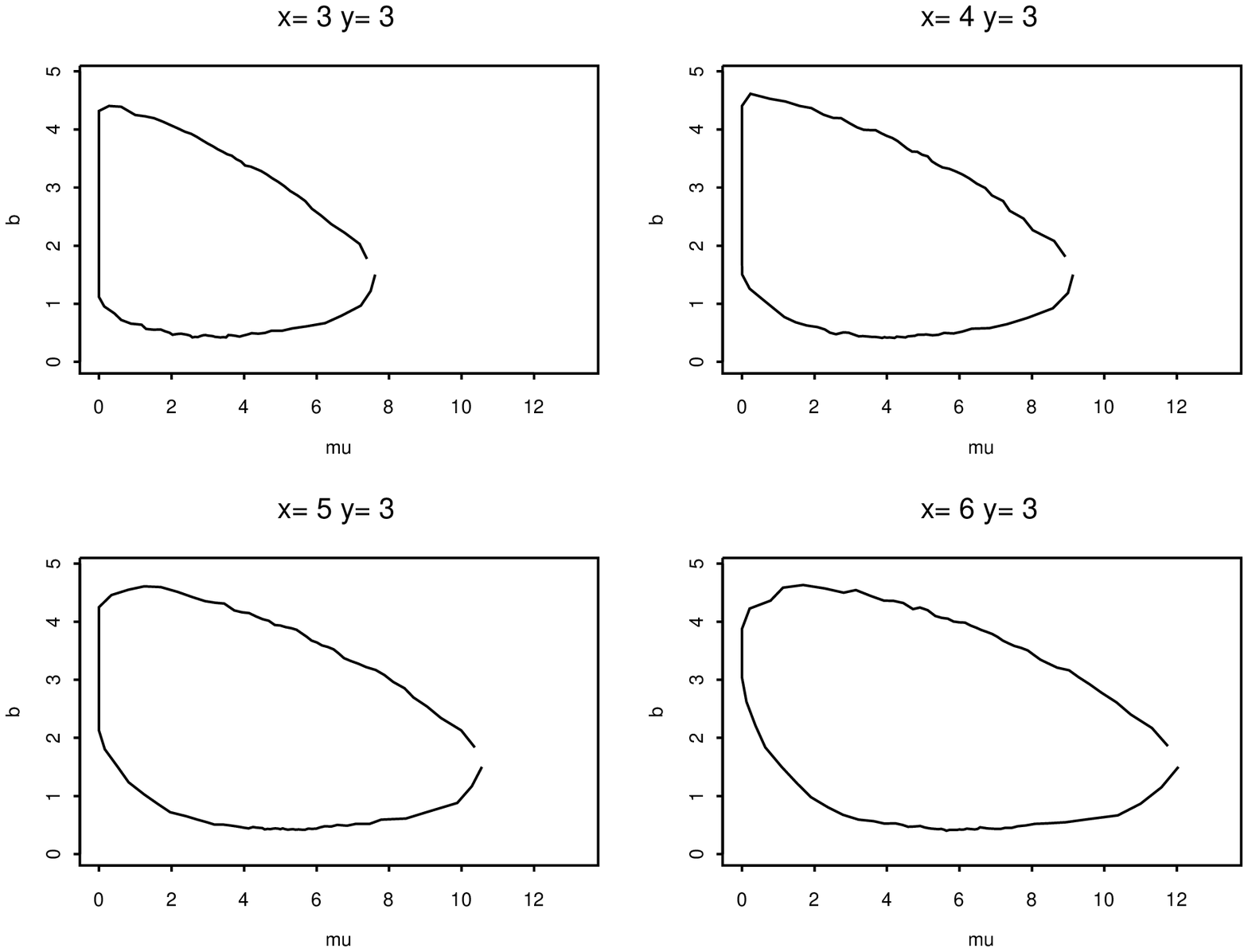,width=403pt,height=316.375pt}
\caption{Two dimensional confidence regions for $x=3,4,5$ and $6$ signal events, and $y=3$ background events. The background region is twice the size of the signal region, and a $90\%$ confidence level was used.}
\end{figure}

\begin{figure}[ht]
\psfig{file=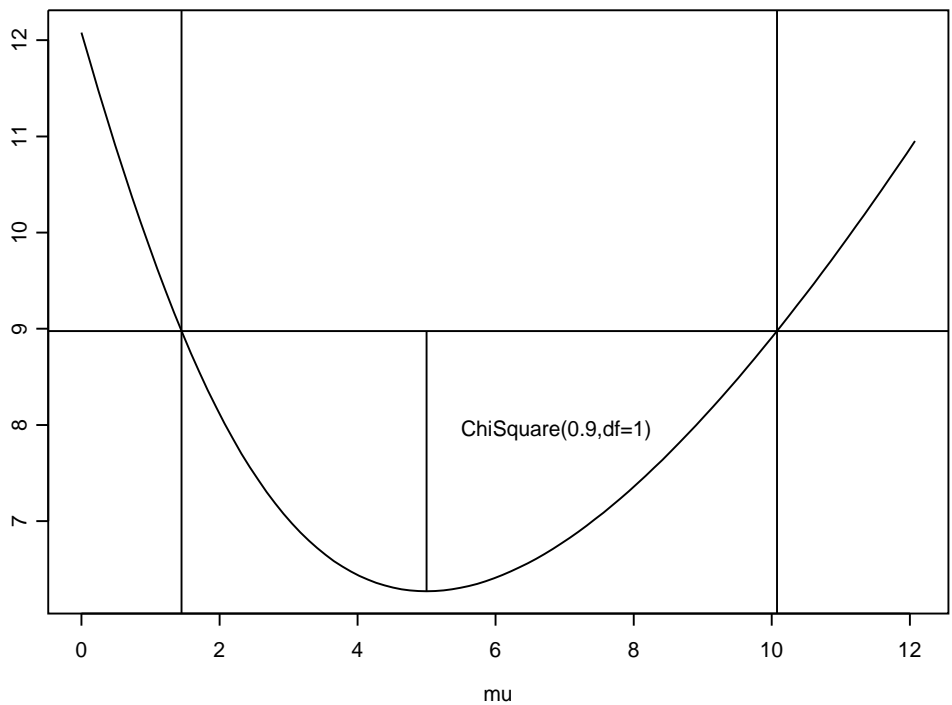,width=403pt,height=316.375pt}
\caption{Profile likelihood function with two-sided limits for the case of $x=6$ events in the signal region and $y=2$ events in the background region. The background region is twice the size of the signal region $(\protect\tau =2)$. The nominal coverage probability is $0.9$.}
\end{figure}

\begin{figure}[ht]
\psfig{file=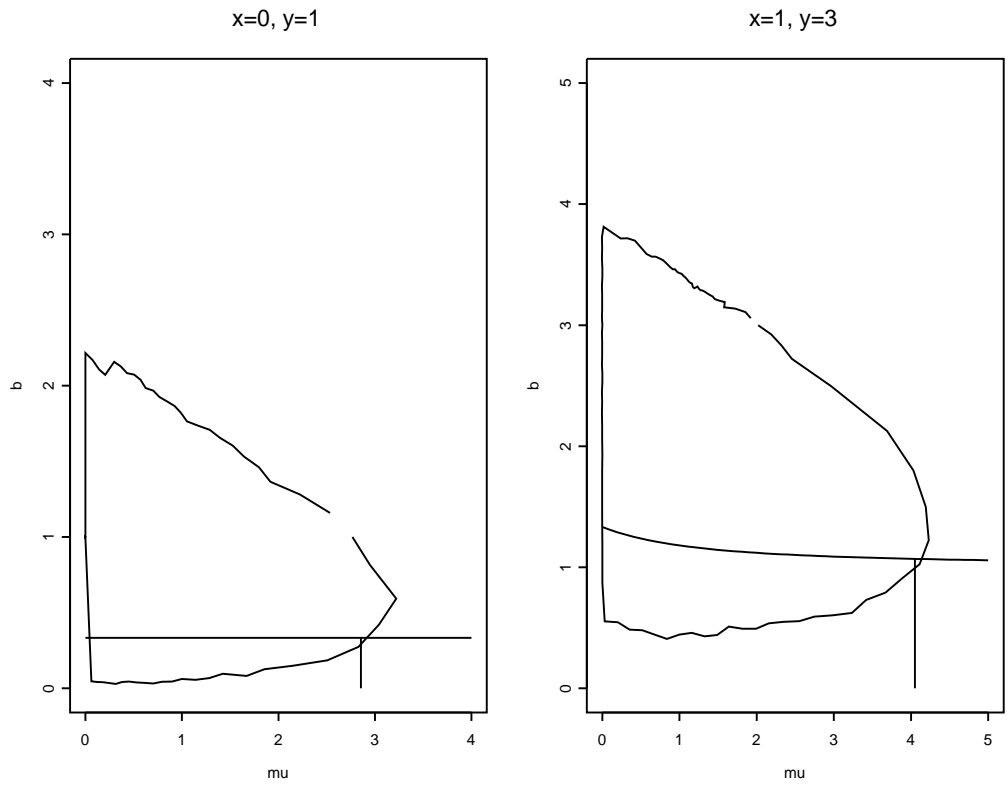,width=403pt,height=316.375pt}
\caption{Confidence region with profile likelihood curve and upper bounds for the cases $x=0,y=1$ and $x=1,y=3$.}
\end{figure}

\begin{figure}[ht]
\psfig{file=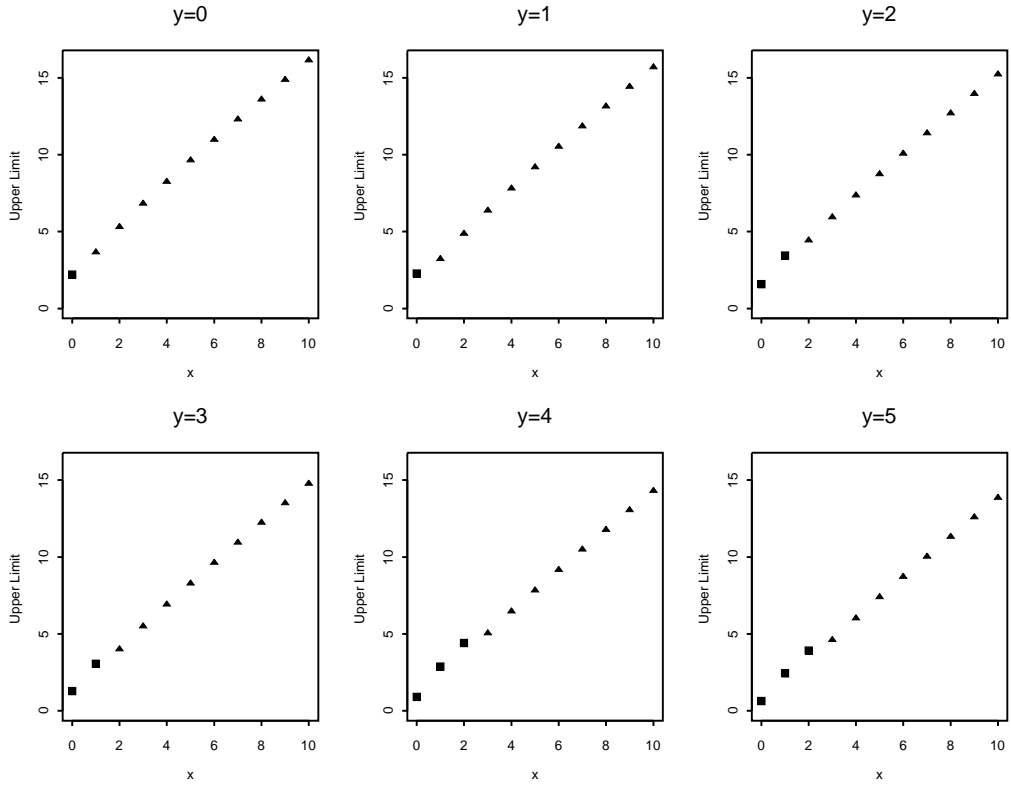,width=403pt,height=316.375pt}
\caption{Upper limits obtained for a range of $x$ and $y$ values. Squares indicate that the confidence region method was used whereas diamonds indicate that the profile likelihood method was used.}
\end{figure}

\begin{figure}[ht]
\psfig{file=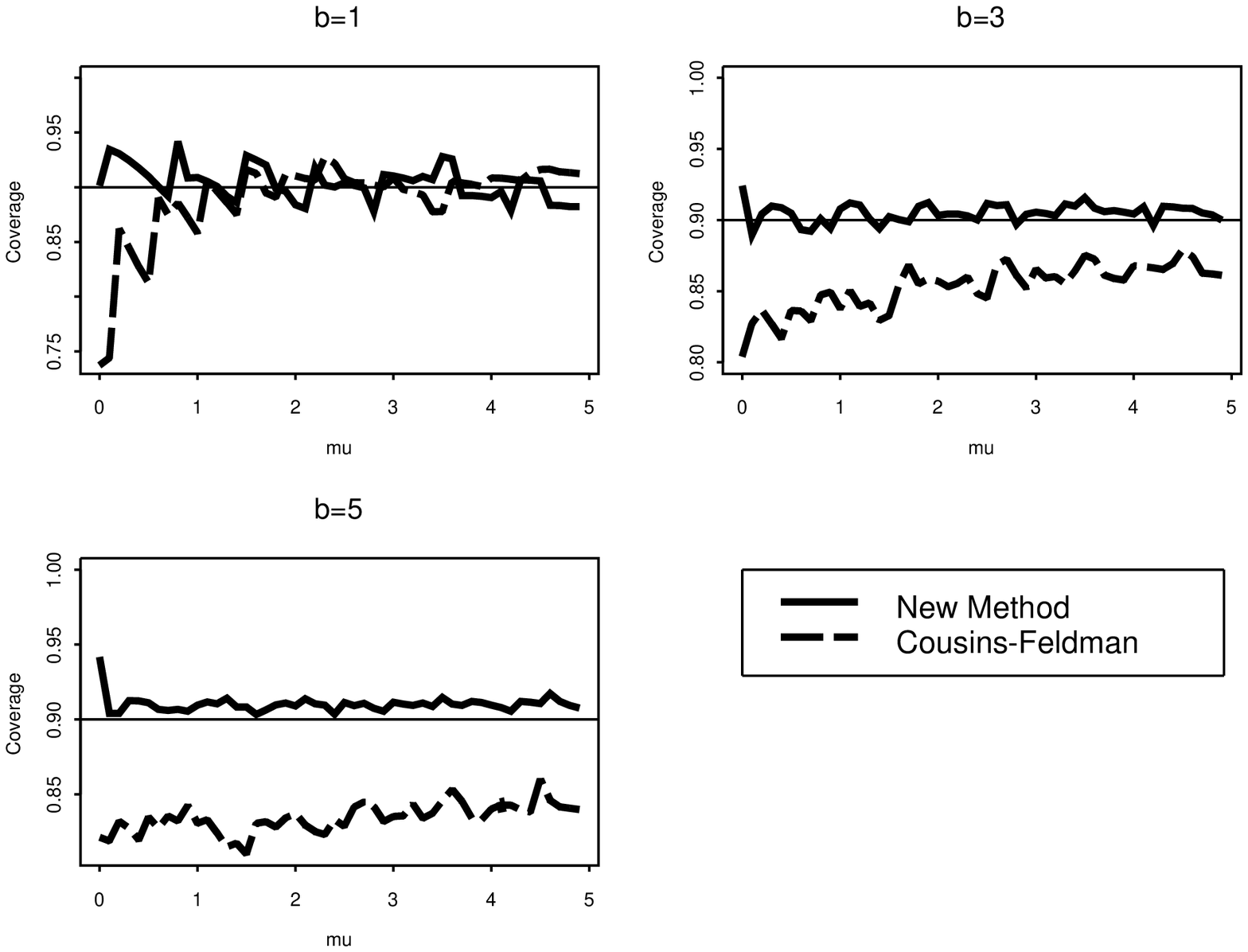,width=403pt,height=316.375pt}
\caption{True coverage rates for our method and for the unified approach of Feldman and Cousins. The background rates are $b=1,3$ and $5$. The signal rates range from $0$ to $5$. We have $\protect\tau =1$ and we use $90\%$ confidence intervals.}
\end{figure}

\begin{figure}[ht]
\psfig{file=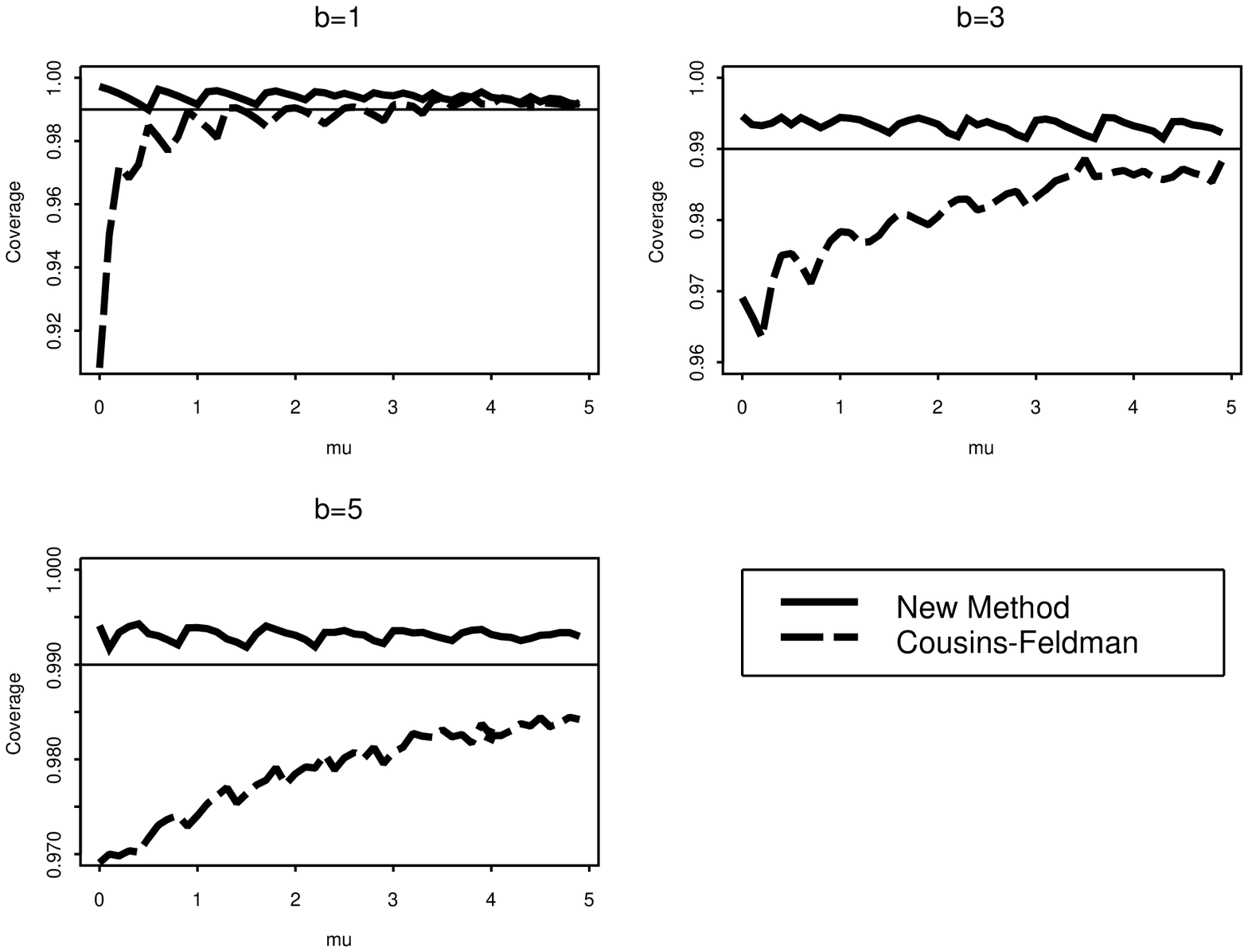,width=403pt,height=316.375pt}
\caption{True coverage rates for our method and for the unified approach of Feldman and Cousins. The background rates are $b=1,3$ and $5$. The signal rates range from $0$ to $5$. We have $\protect\tau =2$ and we use $99\%$ confidence intervals.}
\end{figure}
\begin{figure}[ht]
\psfig{file=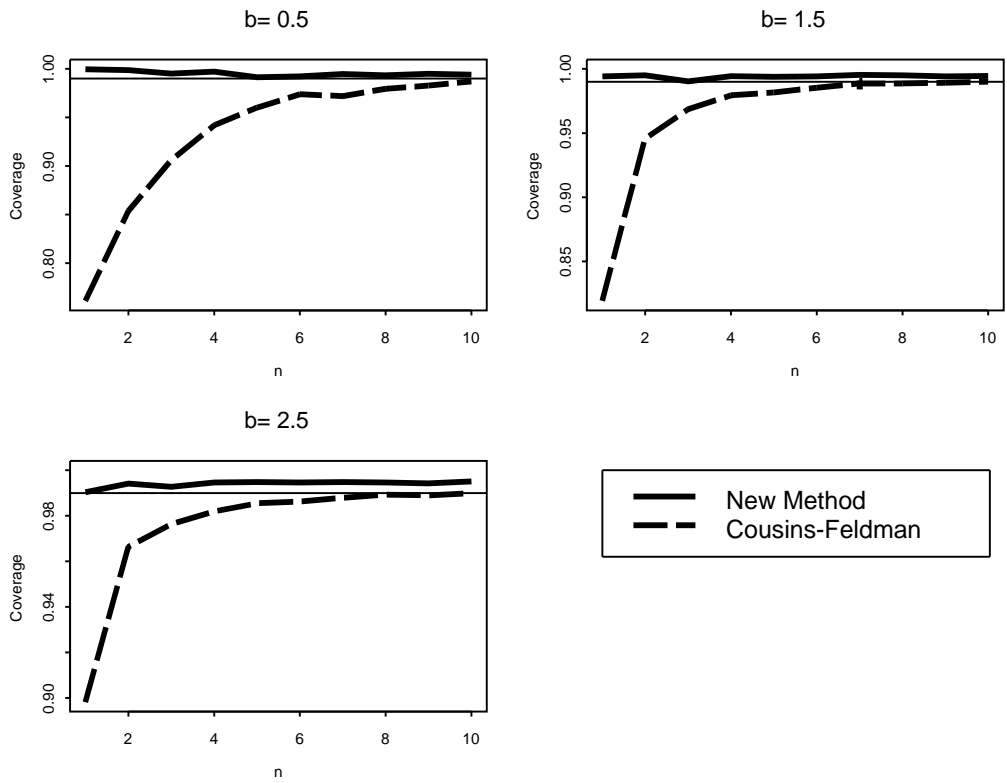,width=403pt,height=315.5625pt}
\caption{Here the background is estimated with Monte Carlo. There is no signal present, and the Monte Carlo was run $n$ times. $99\%$ confidence intervals are used.}
\end{figure}

\begin{figure}[ht]
\psfig{file=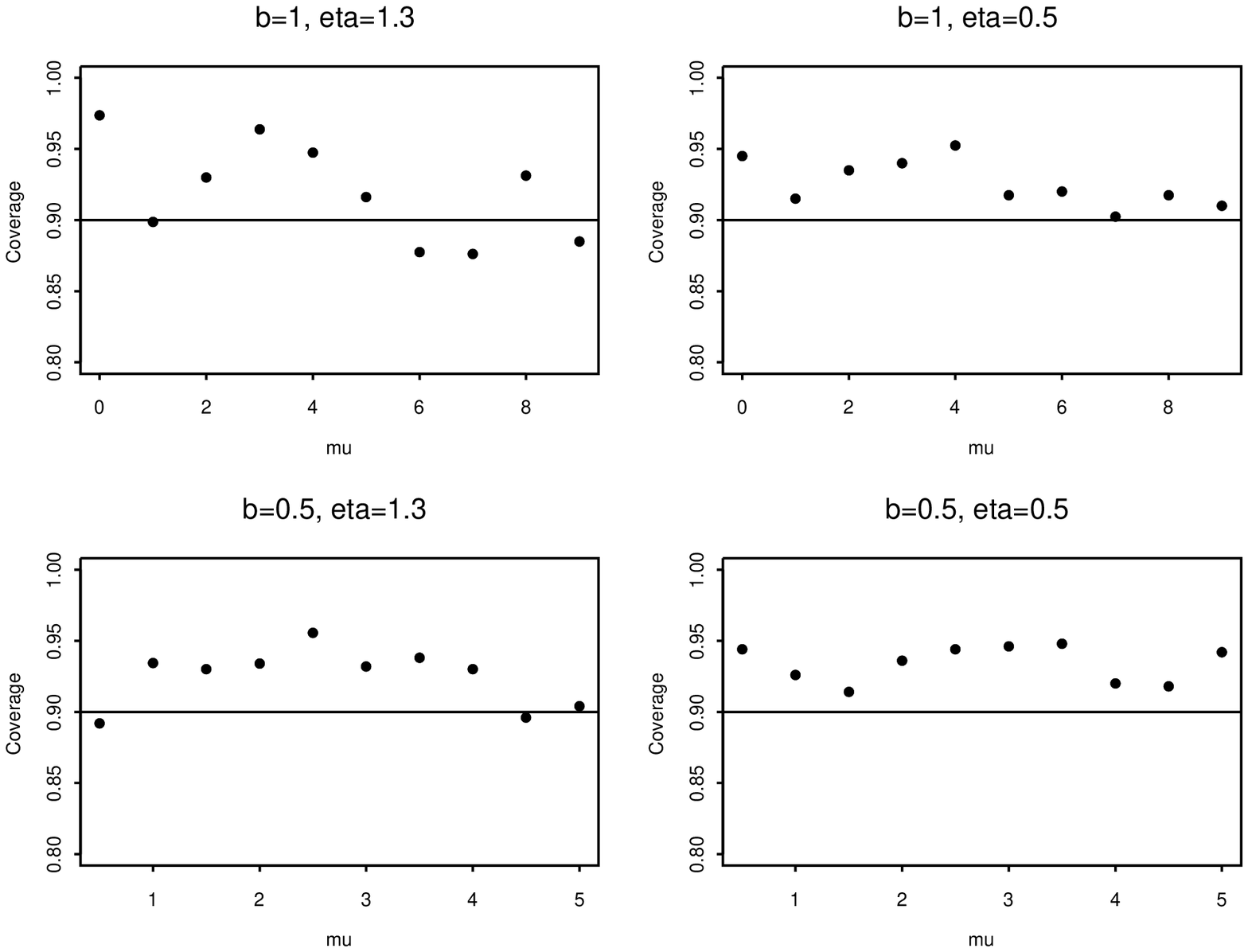,width=403pt,height=316.375pt}
\caption{The case of two different background sources. Coverage rates are based on mini Monte Carlo. The simulation is for the case with a background region twice the size of the signal region. The Monte Carlo was done $10$ times and a $90\%$ confidence interval was found.}
\end{figure}


\begin{thebibliography}{1}
\bibitem[1]{Cousins-Feldman}  R.D. Cousins, G.J. Feldman, ''A Unified
Approach to the Classical Statistical Analysis of Small Signals'', \textit{%
Phys. Rev}, \textbf{D57}, (1998) 3873.

\bibitem[2]{Giunti}  C. Giunti, ''A new ordering principle for the classical
statistical analysis of Poisson processes with background'' , \textit{Phys.
Rev}\emph{\ }\textbf{D59}, 053001 (1999).

\bibitem[3]{Roe-Woodroofe}  B.P. Roe, M.B. Woodroofe, ''Improved Probability
Method for Estimating Signal in the Presence of Background'', \textit{Phys.
Rev} \textbf{D60} 053009 (1999).

\bibitem[4]{Cousins-Highland}  R.D. Cousins, V.L.Highland, ''Incorporating
systematic uncertainties into an upper limit'', \textit{Nucl. Inst. and
Methods A320}, (1992) 331-335.

\bibitem[5]{Casella-Berger}  G. Casella, R.L. Berger, \emph{Statistical
Inference}, Duxburry Press, (1990) 346.

\bibitem[6]{Barlett}  M.S. Bartlett, ''Approximate Confidence Intervals,
Part II: More than one Unknown Parameter'', \textit{Biometrica}\emph{\ }%
\textbf{Vol. 40}, (1953) 306-317.

\bibitem[7]{Lawley}  D.N. Lawley, ''A General Method For Approximating To
The Distribution Of The Likelihood Ratio Criteria'', \textit{Biometrica }%
\textbf{Vol. 43}, (1956) 295-303.

\bibitem[8]{Murphy and Van Der Vaart}  S.A. Murphy, A.W. Van Der Vaart, ''On
Profile Likelihood'', \textit{Journal of the American Statistical Association%
}, \textbf{Vol}. \textbf{95}, (2000), 449-485.

\bibitem[9]{Particle Data Group}  C. Caso \textit{et al.} (Particle Data
Group), Eur. Phys. J. C \textbf{3}, 1 (1998) 177.
\end{thebibliography}
\end{document}